\documentclass{elsart}

\usepackage{graphicx}
\usepackage{epsfig}

\begin{document}

\begin{frontmatter}

\epsfysize3cm
\hspace{-9.5cm}
\epsfbox{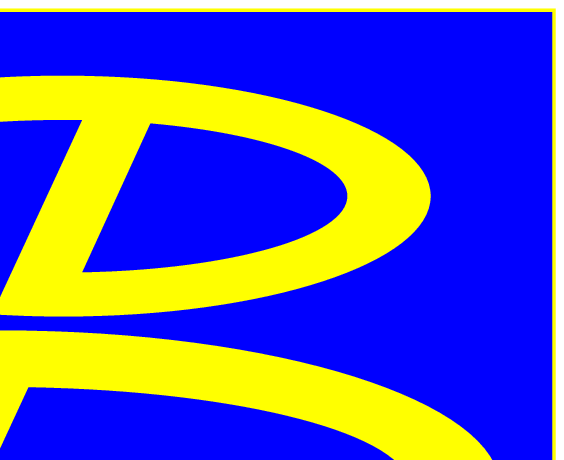}    
\begin{flushright}
\vskip -3cm
\noindent
\hspace*{9.0cm}Belle Preprint 2004-32 \\
\hspace*{9.0cm}KEK Preprint 2004-65 \\
\end{flushright}

\vskip 2cm



\title{
Spectra of prompt electrons from decays of $B^+$ and $B^0$ mesons 
and ratio of inclusive semielectronic branching fractions
}



\collab{Belle Collaboration}
\author[Nagoya]{T.~Okabe}, 
  \author[KEK]{K.~Abe}, 
  \author[TohokuGakuin]{K.~Abe}, 
  \author[KEK]{I.~Adachi}, 
  \author[Tokyo]{H.~Aihara}, 
  \author[Nagoya]{M.~Akatsu}, 
  \author[Tsukuba]{Y.~Asano}, 
  \author[ITEP]{T.~Aushev}, 
  \author[Sydney]{A.~M.~Bakich}, 
  \author[Peking]{Y.~Ban}, 
% ...

% ...
\address[KEK]{High Energy Accelerator Research Organization (KEK), Tsukuba, Japan}
% ...
\address[ITEP]{Institute for Theoretical and Experimental Physics, Moscow, Russia}
% ...
\address[Nagoya]{Nagoya University, Nagoya, Japan}
% ...
\address[Peking]{Peking University, Beijing, PR China}
% ...
\address[Sydney]{University of Sydney, Sydney, NSW, Australia}
% ...
\address[TohokuGakuin]{Tohoku Gakuin University, Tagajo, Japan}
% ...
\address[Tokyo]{Department of Physics, University of Tokyo, Tokyo, Japan}
% ...
\address[Tsukuba]{University of Tsukuba, Tsukuba, Japan}
% ...
\thanks[NovaGorica]{on leave from Nova Gorica Polytechnic, Nova Gorica, Slovenia}


\begin{abstract}
We present spectra of prompt electrons from decays
of neutral and charged $B$ mesons.
The results are based on 140 fb$^{-1}$ of data collected by 
the Belle detector on the $\Upsilon$(4S) resonance
at the KEKB $e^+e^-$ asymmetric collider.
We tag $\Upsilon(4S) \to B \overline{B}$ events
by reconstructing a $B$ meson
in one of several hadronic decay modes;
the semileptonic decay of the other $B$ meson is inferred
from the presence of an identified electron.
We obtain for charged and neutral $B$ mesons the partial rates of
semileptonic decay, to electrons with momentum greater than 0.6\,GeV/$c$
in the $B$ rest frame, and
their ratio $b_+/b_0 =   1.08 \pm 0.05 \pm 0.02$,
where the first and second errors are 
statistical and systematic, respectively.
\end{abstract}

\end{frontmatter}

\section{\boldmath Introduction}
\label{sec:introduction}
The inclusive semileptonic $B$ meson decay branching fraction
${\mathcal{B}}(B \rightarrow X\ell\nu)$ is a fundamental quantity
that is required to fully understand $B$ meson decays.
The decay is believed to be dominated by a spectator process, 
where the $b$ quark is coupled
to a $c$ or $u$ quark and a virtual $W$ boson,
while the accompanying quark in the meson, the so-called spectator,
plays no direct role.
Therefore, the theoretical treatment is relatively simple, and
the semileptonic width $\Gamma_{\rm SL}$ can be readily predicted.
However,
it has long been a puzzle that while theoretical calculations
predict values of ${\mathcal{B}}(B \rightarrow X\ell\nu)$
higher than 12\%~\cite{bib:Puzzle:Bigi},
most measurements have been consistently lower,
at 10--11\%~\cite{bib:PDG2004}.
The discrepancy may be attributed to the uncertainty
in predicting the hadronic decay width $\Gamma_{\rm had}$,
where contributions from non-spectator processes are significant.
The non-spectator contribution depends on
the flavor of the accompanying quark,
while this is not the case for $\Gamma_{\rm SL}$.
Therefore, it may result in
unequal ${\mathcal{B}}(B \rightarrow X\ell\nu)$ values
for neutral and charged $B$ mesons,
hereafter referred to as $b_0$ and $b_+$ respectively.
The ratio $b_{+}/b_{0}$ is equal to
the $B$ lifetime ratio $\tau_{+}/\tau_{0}$
assuming equality in $\Gamma_{\rm SL}$.
The $B$ lifetime ratio is measured well~\cite{bib:PDG2004}.
However, only a few measurements have addressed
$b_{+}$ and $b_{0}$ separately,
and the uncertainties have been large due to low efficiencies
for tagging neutral and charged events~\cite{bib:B+B0:Br}.
The deviations from unity for
both the lifetime ratio and the $b_+/b_0$ ratio
are predicted to be of order 10\%~\cite{bib:Bdecays:Bigi}.

Furthermore, measurement of ${\mathcal B}(B \to X \ell \nu)$ combined
with the lifetime is one of the favored methods to determine
the Cabibbo-Kobayashi-Maskawa matrix element
$|V_{cb}|$~\cite{bib:CKM}.
Heavy-Quark-Expansions (HQEs)~\cite{bib:HQEs} have become
a useful tool to calculate the correction
due to strong interaction effects.
There have been some attempts
to improve the determination of $|V_{cb}|$,
by fitting the perturbative and non-perturbative parameters in HQEs
to the data of the hadronic invariant mass ($M_X$) and 
the lepton energy ($E_{\ell}$) moments 
in the $B$ semileptonic decay~\cite{bib:GlobalFit}.
Recently, the BaBar collaboration performed a fit 
to the partial $B \to X_c e \nu$ branching fraction and
the $M_X$ and $E_{\ell}$ moments,
with varied cutoffs on the lepton energy,
to extract $|V_{cb}|$ and the total $B \to X_c e \nu$
branching fraction as well as
the HQE parameters~\cite{bib:GlobalFit:Babar}
on a consistent basis.

In this paper 
we report measurements of $b_0$ and $b_+$ in the electronic channel
with an electron momentum requirement $p^\ast \geq 0.6$\,GeV/$c$,
as measured in the rest frame of the $B$ meson.
These measurements are based on data collected by
the Belle detector~\cite{bib:BELLE} 
at the KEKB asymmetric $e^+ e^-$ collider~\cite{bib:KEKB}, 
which provides copious production of $B\overline{B}$ meson pairs 
on the $\Upsilon(4S)$ resonance.
In this analysis, one $B$ meson is fully reconstructed in 
one of several hadronic decay modes
to determine its charge, flavor, and momentum, and
is referred to as the tag side $B$  ($B_{\rm tag}$) in the event.
The semileptonic decay of the other $B$ meson,
referred to as the spectrum side $B$ ($B_{\rm spec}$),
is then measured in its rest frame, determined from $B_{\rm tag}$,
without smearing due to the $B$ motion.
Prompt semileptonic decays ($b \to x e \nu$) can be separated 
from secondary decays ($b \to c \to y e \nu$), 
based on the correlation
between the $B_{\rm tag}$ flavor and the electron charge.
To exploit the advantages of this method requires a large sample of
$B\overline{B}$ events because the full reconstruction efficiency is 
rather low, typically of the order of 0.1\%.
Our high integrated luminosity enables us to perform this measurement
with higher accuracy than previously achieved.

\section{\boldmath Data Set and Belle Detector}
\label{sec:datasetandbelledetector}

The results presented in this paper are based on 
a 140 fb${}^{-1}$ data sample
accumulated on the $\Upsilon(4S)$ resonance, 
which contains $152.0 \times 10^6 B \overline{B}$ pairs.
The center-of-mass energy is $\sqrt{s} \simeq 10.58$\,GeV.
An additional 15 fb$^{-1}$ data sample taken at
a center-of-mass energy 60\,MeV 
below the $\Upsilon(4S)$ resonance is used to evaluate 
background from the $e^+ e^- \to q \bar{q}$ ($q=u,d,s,c$) process.
A detailed Monte Carlo (MC) simulation, which fully describes
the detector geometry and response and
is based on GEANT~\cite{bib:GEANT},
is applied to study backgrounds in the $B_{\rm tag}$ reconstruction,
backgrounds in the signal electron detection,
and corrections to the signal selection efficiency due to the tagging.
In the MC simulation, generic $B\overline{B}$ decays are simulated
using the QQ98 generator~\cite{bib:QQ98}.

The Belle detector is a large-solid-angle magnetic spectrometer
that consists of a three-layer silicon vertex detector (SVD),
a 50-layer central drift chamber (CDC), 
an array of aerogel threshold \v{C}erenkov counters (ACC),
a barrel-like arrangement of
time-of-flight scintillation counters (TOF), and
an electromagnetic calorimeter comprised of CsI(Tl) crystals (ECL)
located inside a superconducting solenoid coil 
that provides a 1.5~T magnetic field.  
An iron flux-return located outside of the coil is instrumented
to detect $K_L^0$ mesons and to identify muons (KLM).  
The detector is described in detail elsewhere~\cite{bib:BELLE}.

\section{\boldmath Fully Reconstructed Tagging}
\label{sec:eventselection}
Neutral $B_{\rm tag}$ candidates are reconstructed
in the decay modes
$B^0 \to D^{\ast -} \pi^+$, $D^{\ast -} \rho^+$,
$D^{\ast -} a_1^+$ and
$B^0 \to D^- \pi^+$, $D^- \rho^+$, $D^- a_1^+$.
Charged $B_{\rm tag}$ candidates are reconstructed
in the decay modes
$B^+ \to \overline{D}{}^{\ast 0} \pi^+$,
$\overline{D}{}^{\ast 0} \rho^+$,
$\overline{D}{}^{\ast 0} a_1^+$ and
$B^+ \to \overline{D}{}^0 \pi^+$.
The decay modes $B^+ \to \overline{D}{}^0 \rho^+$ and
$\overline{D}{}^0 a_1^+$ are not used here
because of poor purity due to large combinatorial background.
Inclusion of the charge conjugate decays is implied
throughout this paper.

To suppress the non-$b\bar{b}$ background processes from QED, 
$e^+e^- \to \tau^+\tau^-$, and beam-gas events,
we select hadronic events based on
the charged track multiplicity and total visible energy.
The selection procedure is described
in detail elsewhere~\cite{bib:HadronB}.

Charged particle tracks are reconstructed
from hits in the SVD and CDC. 
They are required to satisfy track quality
based on their impact parameters relative to 
the measured profile of the interaction point (IP profile)
of the two beams,
and good measurements in the SVD in the direction of the beam ($z$).
Charged kaons are identified by combining information
on energy deposit ($dE/dx$) in the CDC,
\v{C}erenkov light yields in the ACC 
and time-of-flight measured by the TOF system.
For the nominal requirement,
the kaon identification efficiency is approximately $88\%$ and
the rate for misidentification of pions as kaons is about $8\%$.
Hadron tracks that are not identified as kaons are treated as pions.
Tracks satisfying the lepton identification criteria
are removed from consideration.

Candidate $\pi^0$ mesons are reconstructed using $\gamma$ pairs 
with an invariant mass
within $\pm 30$\,MeV/$c^2$ of the nominal $\pi^0$ mass.
Each $\gamma$ is required to have a minimum energy deposit of:
$E_{\gamma} \geq 50$\,MeV in the barrel region of the ECL,
defined as $32^{\circ}<\theta_{\gamma}<129^{\circ}$;
$E_{\gamma} \geq 100$\,MeV in the forward endcap region,
defined as $12^{\circ}<\theta_{\gamma}<31^{\circ}$ ;
$E_{\gamma} \geq 150$\,MeV in the backward endcap region,
defined as $131^{\circ}<\theta_{\gamma}<155^{\circ}$,
where $\theta_{\gamma}$ denotes the polar angle of the $\gamma$
with respect to the direction opposite to the positron beam.
$K_S^0$ mesons are reconstructed
using pairs of charged tracks that have 
a well reconstructed vertex that is displaced from the IP and
an invariant mass
within $\pm 7.6$\,MeV/$c^2$ of the known $K_S^0$ mass.
$\rho^+$ and $\rho^0$ candidates are reconstructed 
in the $\pi^+ \pi^0$ and $\pi^+ \pi^-$ decay modes
by requiring their invariant masses to be within $\pm 150$\,MeV/$c^2$ 
of the nominal $\rho$ mass.
The $\rho^+$ candidates are required
to satisfy cos$\theta_{\rho} \geq -0.9$,
where $\theta_{\rho}$ is the helicity angle,
defined as
the angle between an axis anti-parallel to the $B$ flight direction
and the $\pi^+$ flight direction in the $\rho$ rest frame.
$a_1^+$ candidates are formed
from combinations of $\rho^0$ and $\pi^+$ candidates
by requiring that the three tracks form
a good vertex and have an invariant mass
between 0.73\,GeV/$c^2$ and 1.73\,GeV/$c^2$.

$\overline{D}{}^0$ candidates are reconstructed
in the four decay modes
$\overline{D}{}^0 \to K^+ \pi^-$, $K^+ \pi^- \pi^0$, 
$K^+ \pi^+ \pi^- \pi^-$ and $K_S^0 \pi^+ \pi^-$. 
$D^-$ candidates are reconstructed
in the decay mode $D^- \to K^+ \pi^- \pi^-$.
The $D^0$($D^-$) candidates are required
to have an invariant mass within
$\pm 30$ ($12$)\,MeV/$c^2$ of the nominal $D^0$($D^-$) mass. 
$\overline{D}{}^{\ast}$ mesons are reconstructed by pairing 
$\overline{D}{}^0$ candidates with pions, 
$D^{\ast -} \to \overline{D}{}^0 \pi^-$ and
$\overline{D}{}^{\ast 0} \to \overline{D}{}^0 \pi^0$.
The $\overline{D}\pi$ pairs are required to have a mass difference
$\Delta m = m_{\overline{D}\pi} - m_{\overline{D}}$ within
$0.142$\,GeV/$c^2<\Delta m<0.149$\,GeV/$c^2$ for $D^{\ast +}$, and
$0.140$\,GeV/$c^2<\Delta m<0.145$\,GeV/$c^2$ for $D^{\ast 0}$.    
All $D^0$ candidates are used for $B$ reconstruction, 
regardless of whether or not the $D^0$ candidate is used 
to reconstruct a $D^{\ast}$ meson.

The selection of $B$ candidates is
based on the beam-constrained mass,
$M_{\rm bc} = \sqrt{E_{\rm beam}^2 - p_{B}^2}$,
and the energy difference, $\Delta E = E_{B} - E_{\rm beam}$, 
where $E_{\rm beam} \equiv \sqrt{s}/2 \simeq 5.290$\,GeV, 
and $p_B$ and $E_B$ are the momentum and
energy of the reconstructed $B$ 
in the $\Upsilon(4S)$ rest frame, respectively.
The background from jet-like $e^+e^- \to q\bar{q}$ processes
is suppressed by event topology based on 
the normalized second Fox-Wolfram moment ($R_2$)~\cite{bib:R2} and 
the angle between the thrust axis of the $B$ candidate and that of 
the remaining tracks in an event ($\cos \theta_{\rm th}$).
Requirements on $R_2$ and $\cos \theta_{\rm th}$,
as well as the $K/\pi$ selection,
are tuned to suppress the background and
depend on the $B_{\rm tag}$ decay mode.
We select $B_{\rm tag}$ candidates in a signal region defined as 
$5.27$\,GeV/$c^2 \leq M_{\rm bc} \leq 5.29$\,GeV/$c^2$ and
$|\Delta E| \leq 0.05$\,GeV.  

\begin{figure}[tb]
\includegraphics[width=1.00\textwidth]{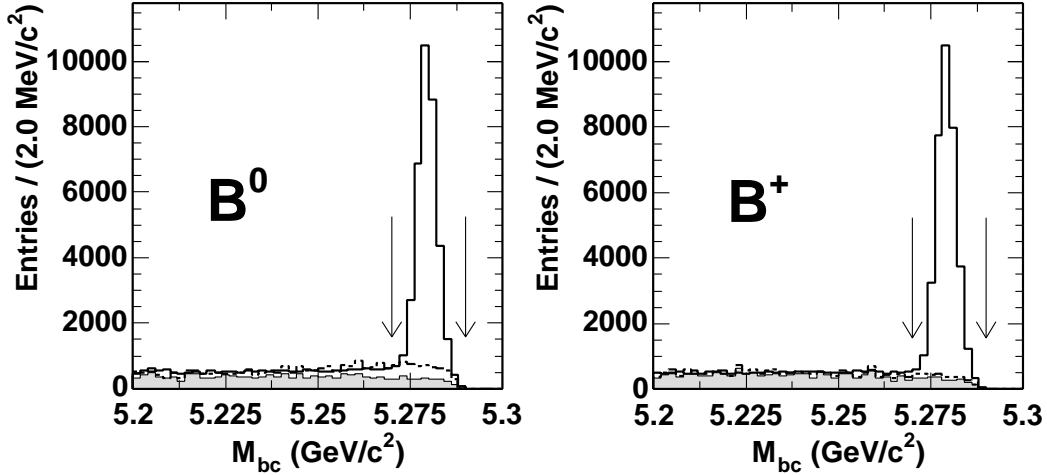}
\caption{
Beam-constrained mass ($M_{\rm bc}$) distributions 
for $B^0$ and $B^+$ candidates
with a $|\Delta E| \leq 0.05$\,GeV requirement. 
The solid histogram corresponds to the on-resonance data.
The hatched histogram is for the off-resonance data
scaled by luminosity.
The dashed histogram indicates the contribution
from the combinatorial background estimated
by scaling the $\Delta E$ sideband
($0.07$\,GeV$\leq \Delta E \leq 0.30$\,GeV).
The arrows indicate the $M_{\rm bc}$ signal region.
}
\label{fig:tagB:Mb}
\end{figure}

Figure~\ref{fig:tagB:Mb} shows the distribution in $M_{\rm bc}$ 
for the neutral and charged $B$ candidates
in the $\Delta E$ signal region.
The $M_{\rm bc}$ signal regions,
indicated by arrows in Fig.~\ref{fig:tagB:Mb}, contain
$36974$ and
$36418$ $B^{0}$ and $B^{+}$ candidates, respectively.
The contribution from the $q\bar{q}$ process is estimated 
by scaling the off-resonance data by the luminosity ratio
with a small correction
due to the energy dependence of the cross section 
and found to be 
$2584 \pm 154$ ($2630 \pm 156$) for $B^0$ ($B^+$) candidates.
The $B$ candidates
remaining after the $q\bar{q}$ background subtraction
contain combinatorial background from $B$ decays,
where some particles are exchanged
between the tag and the spectrum sides (cross-talk).
We estimate the contribution from such combinatorial background to be
$3221 \pm 372$ ($667 \pm 162$) for $B^0$ ($B^+$) candidates,
by scaling the $M_{\rm bc}$ distribution 
in the high $\Delta E$ sideband
($0.07\,\mathrm{GeV} \leq \Delta E \leq 0.30\,\mathrm{GeV}$)
with normalization to the yields 
in the $M_{\rm bc}$ sideband ($M_{\rm bc} \leq 5.26$~GeV/$c^2$)
after the $q\bar{q}$ background subtraction.
Note that we do not apply best $B_{\rm tag}$ candidate selection
for events having multiple candidates,
in order to avoid distorting the $\Delta E$ distribution.
The remaining $31169 \pm 446$ ($33121 \pm 295$) $B^0$ ($B^+$)
candidates, denoted as $N_{\rm tag}(B^0)$ ($N_{\rm tag}(B^+)$),
are from $B\overline{B}$ events,
and are used to normalize the lepton yield
to obtain the semileptonic branching fraction.
These $B_{\rm tag}$ candidates may include a small fraction
with incorrectly assigned $B$ charge and/or flavor
due to particles that are not detected.
The rate of such misassignment is found to be 0.6\%(1.9\%) for events with
(without) electrons on the spectrum side, according to the MC simulation,
and its effect on the determination of $b_+$ and $b_0$ is found to be less  
than 0.1\%.
In order to obtain the electron spectra, presented later, we apply the same 
background subtraction to determine the electron yield for tagged events in 
each electron momentum bin. 

\section{\boldmath Electron Selection and Background Subtraction}
\label{sec:leptonselection}
For events passing our $B_{\rm tag}$ selection,
we search for electrons from semileptonic decays of $B_{\rm spec}$.
The electron momentum ($p^{\ast}$) is measured in the $B$ rest frame, 
which is found using the $B_{\rm tag}$ momentum.
Electrons are divided into two categories, based on the correlation
between the electron charge ($q_{e}$) and
$B_{\rm tag}$ flavor ($Q_{\rm tag}$).
With the assignment $Q_{\rm tag} = +1 (-1)$ 
for $\overline{B}{}^0$ and $B^{-}$ ($B^0$ and $B^{+}$), 
electrons having $q_{e} \times Q_{\rm tag} = +1 (-1)$ are referred 
to as ``right (wrong) sign" electrons.

Electron identification is based on
a combination of $dE/dx$ in the CDC,
the response of the ACC,
shower shape in the ECL and the ratio of energy deposit
in the ECL to the momentum measured by
the tracking system~\cite{bib:EID}.
The electron identification efficiency depends on the track momentum.
Based on the MC simulation, the efficiencies are estimated
to be about 90\% in the momentum region
above 1.2\,GeV/$c$ in the $B$ rest frame, 
where electrons from the prompt $B$ decays dominate.
The rate of pions (kaons) to be mis-identified as electrons
is measured using reconstructed
$K_S^0 \to \pi^+ \pi^-$
($D^{\ast +} \to D^0 \pi^{+}$($D^0 \to K^{-}\pi^{+}$))
and found to be less than 0.2\% for electrons
in the same momentum region.

For the determination of semileptonic branching fractions
we use electron candidates with $p^{\ast} \geq 0.6$\,GeV/$c$.
We demand that electrons be detected
in the barrel region of the associated detector system and
with sufficient transverse energy for a good measurement;
we make requirements 
on the laboratory transverse momenta
with respect to the direction opposite to the positron beam,
$p_{t} \geq 0.6$\,GeV/$c$,
and on the laboratory polar angle,
$35^{\circ} \leq \theta \leq 125^{\circ}$.
Radiative energy loss by electrons is corrected for
by adding back energy found in ECL clusters
within 3 degrees of the reconstructed momentum direction.
Backgrounds from $J/\psi$ decays,
photon conversions in the detector and 
$\pi^0$ Dalitz decays are suppressed by imposing veto conditions;
we calculate invariant masses for each electron candidate 
when combined with opposite charge electrons ($m_{ee}$) 
and with additional photons ($m_{ee\gamma}$),
and reject the electron if $m_{ee}$ lies 
within $\pm 49$\,MeV/$c^2$ of the nominal $J/\psi$ mass, 
$m_{ee}$ is less than 100\,MeV/$c^2$ or 
$m_{ee\gamma}$ is within $\pm 32$\,MeV/$c^2$ of
the nominal $\pi^0$ mass. 

The obtained electron spectra include events
from several background processes.
Table~\ref{tbl:NumberOfLepton} summarizes the number of 
detected electrons and
the contributions from each background source.

Backgrounds from $J/\psi$ decays, photon conversion and 
$\pi^0$ Dalitz decays are small after the veto.
The remaining backgrounds,
where one of the pair has escaped detection,
are estimated by the MC simulation.
The background from these processes amounts to 2.0\%
of the yield in the signal region.
The uncertainties are evaluated from the error on each rate.

Contributions from secondary electrons from $B$ decays are modeled by the
MC simulation based on Ref.~\cite{bib:QQ98}
and branching fractions quoted in Ref.~\cite{bib:PDG2004}. 
These include leptons from $\tau$ decays in processes
such as $B \to X \tau^+ \nu$ and $B \to D_s X$ followed by $D_s \to \tau^+ \nu$. 
The uncertainty of their contribution is estimated based on 
${\mathcal B}(b \to \tau \nu + anything) = (2.48 \pm 0.26)$\% in Ref.~\cite{bib:PDG2004}.
Another major source of secondary electrons is the $W^+ \to c\bar{s}
(c\bar{d})$ processes (``upper vertex'' charm) such as $B \to D_s X,
D_s \to Y \ell^+ \nu$ and $B \to D^{(*)}D^{(*)}K^{(*)}, D \to Y \ell^+ \nu$
via $\bar{b} \to \bar{c} c \bar{s}$ and a small contribution from 
$B \to D^{(*)}D^{(*)}$.
The uncertainty of their contribution is estimated based on 
${\mathcal B}(\bar{b} \to \bar{c} c \bar{s}) = (22 \pm 4)$\% in Ref.~\cite{bib:PDG2004}.
The backgrounds from these processes account for 4.3\%
of the yield in the signal region.
The uncertainties are evaluated 
from the errors on the associated branching fractions.

Contributions from misidentified hadrons are estimated
by multiplying the measured fake rates by
the number of additional hadron tracks 
in events containing selected $B_{\rm tag}$.
Here, the hadrons are obtained 
by imposing a lepton identification veto on charged tracks. 
Misidentified hadrons are distributed mainly in the momentum region below 1.5GeV/$c$,
and amount to 0.4\%(1.0\%)
over the whole momentum range of the right(wrong) sign spectra.
 
\begin{table}[tb]
\caption{
Summary of electron yields and estimated backgrounds. 
}
\label{tbl:NumberOfLepton}
\begin{center}
\begin{tabular}{lcccc}
\hline
                                 & \multicolumn{2}{c}{$B^0$ candidate}    & \multicolumn{2}{c}{$B^+$ candidate}  \\
                               &e : right-sign&e : wrong-sign&e : right-sign&e : wrong-sign\\
\hline \hline
on-resonance data                & 2007.0 $\pm$  44.8 &  967.0 $\pm$  31.1  & 2520.0 $\pm$ 50.2 & 450.0 $\pm$ 21.2  \\
\hline                                                                     
scaled off resonance             &    9.2 $\pm$   9.2 &    0.0 $\pm$   9.2  &    9.2 $\pm$  9.2 &   0.0 $\pm$  9.2  \\
estimated combinatorial          &   78.6 $\pm$  10.5 &   33.9 $\pm$   5.5  &   31.9 $\pm$  6.9 &  12.7 $\pm$  3.3  \\
\hline                                                                     
estimated background             &  130.7 $\pm$   1.3 &   73.9 $\pm$   1.1  &  154.3 $\pm$  1.0 &  46.6 $\pm$  0.5  \\
\hspace{3mm}from $J/\psi$        &    2.8 $\pm$   0.1 &    2.9 $\pm$   0.1  &    3.0 $\pm$  0.1 &   3.0 $\pm$  0.1  \\
\hspace{3mm}from Dalitz or conv. &   25.7 $\pm$   0.4 &   25.9 $\pm$   0.4  &   27.7 $\pm$  0.4 &  27.5 $\pm$  0.4  \\
\hspace{3mm}from $\tau$          &   42.7 $\pm$   0.5 &   10.1 $\pm$   0.2  &   52.6 $\pm$  0.6 &   0.8 $\pm$  0.1  \\
\hspace{3mm}from upper vertex    &   50.9 $\pm$   0.5 &   27.3 $\pm$   0.4  &   60.5 $\pm$  0.6 &   9.5 $\pm$  0.2  \\
\hspace{3mm}hadron fakes         &    8.6 $\pm$   1.0 &    7.7 $\pm$   0.9  &   10.5 $\pm$  0.3 &   5.8 $\pm$  0.2  \\
\hline                                                                     
bkg. subtracted                  & 1788.5 $\pm$  47.4 &  859.2 $\pm$  34.0  & 2324.4 $\pm$ 51.4 & 390.7 $\pm$ 23.3  \\
\hline                                                                     
after mixing corr.               & 2063.7 $\pm$  62.3 &  584.0 $\pm$  46.3  & \multicolumn{2}{c}{}                 \\
\hline
after eff. corr.                 & 3298.6 $\pm$ 104.2 & 1067.3 $\pm$  80.8  & 3736.7 $\pm$ 84.0 & 713.5 $\pm$ 42.6  \\
\hline
\end{tabular}
\end{center}
\end{table}

\section{\boldmath Semileptonic Decay Spectra}
\label{sec:semileptonicbranchingfraction}
The spectra after the above background subtraction contain 
electrons from prompt semileptonic $B$ decays and 
from secondary semileptonic charm decays (``lower vertex'' charm).
After background subtraction, 
the number of right(wrong) sign electrons is
$1789\pm47$ ($859\pm34$) for events tagged with $B^0$, and
$2324\pm51$ ($391\pm23$) for those tagged with $B^+$.
For events tagged with a $B^+$, 
electrons with the right and wrong signs correspond 
to those from prompt $B$ and secondary charm decays, respectively.
For events tagged with a $B^0$, 
the effect of $B^0$-$\overline{B}{}^0$ mixing is taken into account, 
by solving the following equations for $N_{\rm p}$ and $N_{\rm s}$,
the number of electrons
from prompt and secondary semileptonic decays, respectively:
\begin{eqnarray}
N_{\rm right} &=& N_{\rm p}(1 - \chi_{d}) + N_{\rm s} \chi_{d}  \nonumber \\
N_{\rm wrong} &=& N_{\rm p} \chi_{d} + N_{\rm s} (1 - \chi_{d}),\nonumber
\end{eqnarray}
where $N_{\rm right}$ and $N_{\rm wrong}$ are
the numbers of right- and wrong-sign electrons 
and $\chi_{d} = 0.186 \pm 0.004$~\cite{bib:PDG2004} is 
the $B^0$-$\overline{B}{}^0$ mixing probability.  

\begin{figure}[tb]
\includegraphics[width=1.00\textwidth]{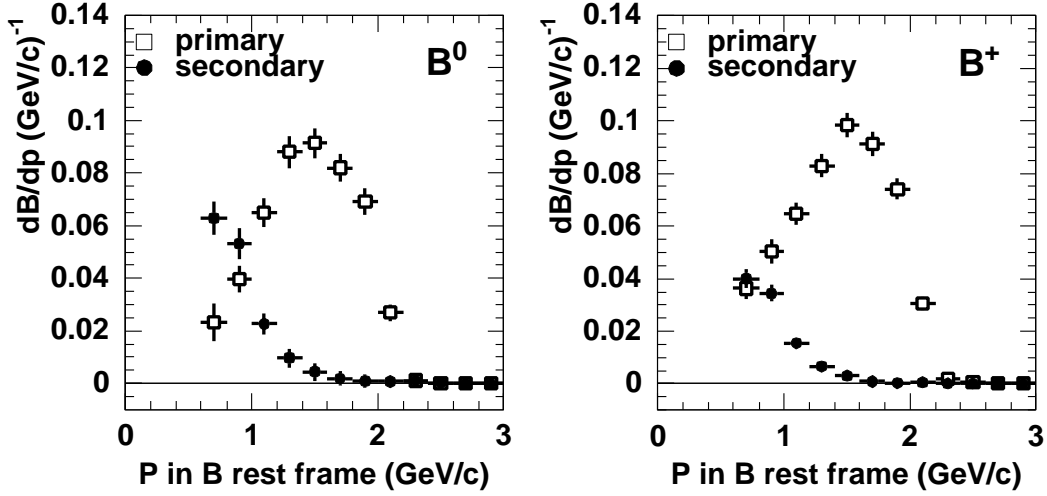}
\caption{
Momentum spectra in the $B$ meson rest frame of electrons
from prompt semileptonic $B$ decay (open squares) and
secondary semileptonic charm decay (closed circles)
for $B^0$ and $B^+$ tagged events. 
The vertical scale shows
the differential branching fraction ($d{\mathcal B}/dp$).
}
\label{fig:lepton:spectrum}
\end{figure}

The electron detection efficiency is corrected
for detector acceptance,
tracking and electron selection efficiencies,
where the correction is evaluated with the MC simulation.
We also take into account a correlation due to the difference
in the event tagging efficiency between events
where the $B_{\rm spec}$ decays semileptonically and
those where it decays hadronically.
This effect is referred to hereafter as ``tag bias''.
In the MC simulation, it is found that
the difference depends on track multiplicity in the event,
which alters the detection efficiency of charged,
$\pi^0$ and $K_S^0$ particles
used for the $B_{\rm tag}$ reconstruction.
The tag bias effect is estimated from the change of semileptonic
decay fraction in the tagged sample using the generator information
in the MC simulation, which is 8\%(6\%) for $B^0$($B^+$).
Figure~\ref{fig:lepton:spectrum} shows the $p^{\ast}$ spectra
from the prompt $B$ and the secondary semileptonic decays 
obtained separately for $B^0$ and $B^+$.
The differential branching fractions $d{\mathcal B}/dp$ are obtained
from the number of electrons,
normalized by $N_{\rm tag}(B^0)$ or $N_{\rm tag}(B^+)$.
Table~\ref{tbl:diff-br} shows
obtained differential branching fraction for each bin.
Both in Figure~\ref{fig:lepton:spectrum} and Table~\ref{tbl:diff-br},
the errors are statistical only.
The analysis of systematic uncertainties presented in detail
in Section~\ref{sec:systematicerror} shows that
they are momentum independent and
the common systematic error of 3.4\% and 3.6\% can be ascribed
to all the bins for $B^0$ and $B^+$, respectively.

The partial branching fractions for the electron channel, 
integrated over the momentum region above 0.6\,GeV/$c$, are
$b_0$($p^{\ast} \geq 0.6$\,GeV/$c$) = ($ 9.83 \pm 0.34$)\% and
$b_+$($p^{\ast} \geq 0.6$\,GeV/$c$) = ($10.62 \pm 0.25$)\% for 
$B^0$ and $B^+$, respectively.
Their average and ratio are found to be
$b$($p^{\ast} \geq 0.6$\,GeV/$c$) = ($ 10.34 \pm 0.20$)\%, and
$b_+/b_0$($p^{\ast} \geq 0.6$\,GeV/$c$) = $1.08 \pm 0.05$.

\begin{table}[tb]
\caption{
Differential branching fractions of $B^0$ and $B^+$ for each bin.
The last column shows their ratio.
The errors are statistical only.
The common systematic error of 3.4\%, 3.6\% and 1.9\% can be ascribed
to all the bins for $B^0$, $B^+$ and $B^+/B^0$, respectively.
}
\label{tbl:diff-br}
\begin{center}
\begin{tabular}{cccc}
\hline
 & \multicolumn{2}{c}{$d{\mathcal B}/dp~~~($GeV/$c){}^{-1}$} & \\
p(GeV/$c$) & $B^0$ & $B^+$ & $ B^{+}/B^{0} $ \\ \hline
\hline
 0.6--0.8  &  $ 0.0234 \pm 0.0071 $  &  $ 0.0362 \pm 0.0040 $  & $ 1.547 \pm 0.502 $ \\
 0.8--1.0  &  $ 0.0401 \pm 0.0053 $  &  $ 0.0503 \pm 0.0047 $  & $ 1.253 \pm 0.203 $ \\
 1.0--1.2  &  $ 0.0656 \pm 0.0056 $  &  $ 0.0648 \pm 0.0042 $  & $ 0.987 \pm 0.106 $ \\
 1.2--1.4  &  $ 0.0889 \pm 0.0061 $  &  $ 0.0831 \pm 0.0045 $  & $ 0.934 \pm 0.082 $ \\
 1.4--1.6  &  $ 0.0925 \pm 0.0058 $  &  $ 0.0985 \pm 0.0048 $  & $ 1.065 \pm 0.085 $ \\
 1.6--1.8  &  $ 0.0829 \pm 0.0055 $  &  $ 0.0913 \pm 0.0046 $  & $ 1.101 \pm 0.091 $ \\
 1.8--2.0  &  $ 0.0700 \pm 0.0050 $  &  $ 0.0742 \pm 0.0041 $  & $ 1.060 \pm 0.096 $ \\
 2.0--2.2  &  $ 0.0271 \pm 0.0031 $  &  $ 0.0304 \pm 0.0026 $  & $ 1.121 \pm 0.160 $ \\
 2.2--2.4  &  $ 0.0010 \pm 0.0007 $  &  $ 0.0019 \pm 0.0007 $  & $ 1.892 \pm 1.408 $ \\
 2.4--2.6  &  $ 0.0001 \pm 0.0002 $  &  $ 0.0005 \pm 0.0003 $  & $ 3.912 \pm 5.849 $ \\
\hline
\end{tabular}
\end{center}
\end{table}

\section{\boldmath Systematic Error}
\label{sec:systematicerror}
The systematic uncertainties
on the partial semileptonic branching fractions are 
evaluated separately for $b_0$ and $b_+$, 
and are summarized in Table~\ref{tbl:sys1}.

The uncertainty in $N_{\rm tag}$ is associated mainly with 
the procedure for the combinatorial background subtraction
described earlier.
After applying the same procedure on the simulated data and
comparing the number of obtained tagged candidates with the true
number, the systematic uncertainty due to $N_{\rm tag}$
determination is estimated to be 1.0--1.9\%.

The uncertainty due to the tag bias correction
is estimated to be 1.3\% from the uncertainties of
the charged particle and photon multiplicity
dependence in the $B_{\rm tag}$ reconstruction.
We find multiplicity difference between data and the simulation
to be about 0.1 for charged particles and 0.2 for photons.
These differences propagate to the reconstruction efficiencies of
charged particles, $\pi^0$ and $K_S^0$, and hence
the $B_{\rm tag}$ which is reconstructed based on
8.1 charged particles, 2.7 $\pi^0$ and
1.1 $K_S^0$ per event on average.
We add another 1.8\%(1.6\%) uncertainty due to the statistics in the MC
simulation to determine the tag bias correction factor for $B^0$($B^+$).

The uncertainty on the tracking efficiency
is determined based on a study using
$D^{\ast +} \to D^0(\to K_S^0 \pi^+\pi^-) \pi^{+}$ decays.
In this study,
the yield of fully reconstructed $D^{\ast}$ mesons is
to be compared to that using partial reconstruction,
where one pion from $K_S^0$ is not used.
A $\pm 1.0$\% uncertainty is assigned for the tracking efficiency
by taking the difference of the yield ratio
between the experimental data and the MC simulation.

The uncertainty on the electron identification efficiency is 
one of the largest sources of systematic error.
It is estimated to be $\pm 2.1$\% from the difference
between the efficiency 
determined from the MC simulation and that based on a sample of 
simulated tracks embedded in beam data.
The uncertainty on the fake electron rate is studied
by comparing the fake rates measured with
$K_S^0 \to \pi^{+}\pi^{-}$ decays
in real data and in the MC simulation.
The uncertainty on $b_0$ and $b_+$ is estimated 
to be $\pm 0.1$\%.

The uncertainties on the background subtractions
from $J/\psi$ decays, converted electrons, 
$\tau$ decays, and the ``upper vertex'' processes
are evaluated from the error on each rate, as described above.
The uncertainty in $b_0$ and $b_+$ for the ``upper vertex'' 
processes is ($0.5 - 0.6$)\%.

The uncertainty from the mixing probability $\chi_d$ is determined 
based on its quoted error in Ref.~\cite{bib:PDG2004}, 
and contributes $\pm 0.4$\% to the systematic error on $b_0$.

The uncertainty in the continuum subtraction is attributed to the 
normalization between on- and off-resonance data, and is estimated 
to be $\pm 0.1$\%
based on the error of the relative luminosity measurement.

The overall systematic errors are evaluated 
by adding these errors in quadrature.
The systematic error on the ratio $b_{+}/b_{0}$ is small 
because several sources of systematic error cancel in the ratio.
The remaining sources of systematic error are mainly 
$N$($B_{\rm tag}$) estimation ($1.9$\%) and mixing ($0.4$\%). 
The overall systematic errors on the partial branching fractions are
3.6\% for $b_{+}$, 3.4\% for $b_{0}$
and 1.9\% for $b_{+}/b_{0}$.
%
\begin{table}[tb]
\caption{Contributions to the systematic error.}
\label{tbl:sys1}
\begin{center}
\begin{tabular}{|l||c|c|c|}
\hline 
Source & 
$\Delta b_{0}   / b_{0}$ (\%)   & $\Delta b_{+}   / b_{+}$ (\%) &
$\Delta \frac{b_+}{b_0}/\frac{b_+}{b_0}$ (\%) \\
\hline \hline
$N_{\rm tag}$            & $\pm 1.0$ & $\pm 1.9$          & $\pm 1.9$ \\ \hline
Tag-bias                 & $\pm 2.2$ & $\pm 2.1$          & ---       \\ \hline
Tracking                 & \multicolumn{2}{c|}{$\pm 1.0$} & ---       \\ \hline
PID efficiency           & \multicolumn{2}{c|}{$\pm 2.1$} & $ <  0.1$ \\ \hline
Hadron fakes             & \multicolumn{2}{c|}{$\pm 0.1$} & $ <  0.1$ \\ \hline
$e$ from $J/\psi$        & \multicolumn{2}{c|}{$ <  0.1$} & $ <  0.1$ \\ \hline
$e$ from conversion      & \multicolumn{2}{c|}{$ <  0.1$} & $ <  0.1$ \\ \hline
$e$ from $\tau$          & \multicolumn{2}{c|}{$\pm 0.3$} & $ <  0.1$ \\ \hline
$e$ from upper vertex    & $\pm 0.6$ & $\pm 0.5$          & $ <  0.1$ \\ \hline
Mixing                   & $\pm 0.4$ & ---                & $\pm 0.4$ \\ \hline
Continuum subtraction    & \multicolumn{2}{c|}{$\pm 0.1$} & $\pm 0.1$ \\ \hline
\hline
Total                    & $\pm 3.4$ & $\pm 3.6$          & $\pm 1.9$ \\ \hline
\end{tabular}
\end{center}
\end{table}

\section{\boldmath Results and Summary}
\label{sec:resultsandsummary}
Including the above systematic errors,
the partial semileptonic branching fractions are
\begin{eqnarray}
b_0 ( p^{\ast} \geq 0.6 \mathrm{\,GeV}/c) & = & (  9.83 \pm 0.34 \pm 0.33 )\%      \nonumber \\
b_+ ( p^{\ast} \geq 0.6 \mathrm{\,GeV}/c) & = & ( 10.62 \pm 0.25 \pm 0.39 )\%      \nonumber
\end{eqnarray}
and their average and ratio are found to be
\begin{eqnarray}
b   ( p^{\ast} \geq 0.6 \mathrm{\,GeV}/c) & = & ( 10.34 \pm 0.20 \pm 0.36 )\%    \nonumber \\
b_+/b_0 ( p^{\ast} \geq 0.6 \mathrm{\,GeV}/c) & = & 1.08 \pm 0.05 \pm 0.02.  \nonumber
\end{eqnarray}
These average values are calculated with weights 
determined by the statistical error of each subsample.

The average partial branching fraction $b$ is consistent with
our previous measurement~\cite{bib:dilepton:BELLE},
with the overall error improved by 15\%,
and it is also consistent with recent measurements
on the $\Upsilon(4S)$
resonance by BaBar and CLEO~\cite{bib:Partial:Br}
with the same minimum momentum requirement.
Our results are the most precise separate determinations
of $b_{+}$, $b_{0}$ and their ratio $b_{+}/b_{0}$.
The observed $b_{+}/b_{0}$ ratio is consistent with
the $B^+/B^0$ lifetime ratio
$\tau_{+}/\tau_{0} = 1.086 \pm 0.017$~\cite{bib:PDG2004}.
Furthermore, as shown in Table~\ref{tbl:diff-br},
the ratio of the differential branching fraction for each momentum bin
is consistent with $\tau_{+}/\tau_{0}$.
There is no indication that the na\"{\i}ve expectation of
equal $\Gamma_{\rm SL}$
for charged and neutral $B$ mesons break down
in the measured range of electron momentum.

The present analysis method using fully reconstructed tags
can be extended to a separate determination of
the $M_X$ and $E_{\ell}$ moments
in the $B^+$ and $B^0$ semileptonic decays.
The partial branching fractions obtained in
the present work can be used
as part of a combined fit of HQE parameters
to the full set of the moments
to determine the total branching fraction as well as $|V_{cb}|$.
In contrast to measurements
based on samples with $B^+/B^0$ admixtures,
such an approach will help to eliminate
the uncertainty in $|V_{cb}|$
due to the production ratio of $B^+$ and $B^0$
on the $\Upsilon(4S)$ resonance ($f_+/f_0$),
and will also provide a useful cross check of assumptions
behind the HQE theory, such as quark-hadron duality.
These extensions to this analysis
will be reported in future articles.

\ack
We thank the KEKB group for the excellent operation of the
accelerator, the KEK Cryogenics group for the efficient
operation of the solenoid, and the KEK computer group and
the National Institute of Informatics for valuable computing
and Super-SINET network support. We acknowledge support from
the Ministry of Education, Culture, Sports, Science, and
Technology of Japan and the Japan Society for the Promotion
of Science; the Australian Research Council and the
Australian Department of Education, Science and Training;
the National Science Foundation of China under contract
No.~10175071; the Department of Science and Technology of
India; the BK21 program of the Ministry of Education of
Korea and the CHEP SRC program of the Korea Science and
Engineering Foundation; the Polish State Committee for
Scientific Research under contract No.~2P03B 01324; the
Ministry of Science and Technology of the Russian
Federation; the Ministry of Education, Science and Sport of
the Republic of Slovenia;
the Swiss National Science Foundation;
the National Science Council and
the Ministry of Education of Taiwan; and the U.S.\
Department of Energy.




\end{document}